\documentclass{article}
\usepackage{amsmath}
\usepackage[latin9]{inputenc}
\usepackage{caption}
\usepackage[american]{babel}
\usepackage[final]{pdfpages}
\usepackage{xcolor}
\usepackage{titlesec}
\usepackage{float}
\usepackage[numbers]{natbib}
\usepackage{rotating}
\usepackage{subfig}
\usepackage[labelfont=bf]{caption}
\usepackage[width=.75\textwidth]{caption}
\makeatletter
\providecommand{\tabularnewline}{\\}
\begin{document}

\title{Shell Model Symmetries}

\author{Levering Wolfe and Larry Zamick\\
 Department of Physics and Astronomy, \\
 Rutgers University, Piscataway, New Jersey 08854}
 
\maketitle
\begin{abstract}
We discuss L=0 vs L=2 couplings, symmetries of the pairing interaction
for neutrons and protons, and the J$_{max}$ interaction. We show
that certain schematic interactions yield exponentially decreasing
transition strengths.We compare shell model calculations  of B(E2)'s and quadrupole moments in the p-f shell  with collective model results. 
\end{abstract}

It is a pleasure to be able to contribute to proceedings honoring
this milestone of Franco Iachello's career. We will start with what
may be regarded as early elementary precursors to his seminal papers
with Akito Arima and others {[}1,2{]} on the interaction boson approximation
-IBA and IBM. Then, as a change of pace we will discuss most recent
work where we focus on matrix models of strength distributions for
which there are many problems that are not yet understood.

\section{MBZE WAVE FUNCTIONS}

We first discuss single j shell calculations in the f$_{7/2}$ shell
which were carried out around 1964 byMcCullen, Bayman and Zamick {[}3{]}
, Ginocchio and French {[}4{]} , and with improved input (MBZE) in
2006 {[}5{]}.As our input Hamiltonian we take matrix elements from
experiment--from the spectrum of $^{42}$Sc. The T=1matrix elements
for J=0,2,4,6 (in MeV) are 0.000, 1.5803,2.8153, and 3.2420. The T=0
ones for J=1,3,5,7 are 0.6111,1.4903,1.5161 and 0.6163.

In Table 1 we show wave functions of $^{44}$Ti in the single j shell
model. The columns are amplitudes for the protons to couple to J$_{p}$
and neutrons to J$_{n}$ for a state of total angular momentum J=0.

\begin{center}

\captionof{table}{Wave functions of the lowest J= 0$^{+}T=0$ state in $^{44}$Ti
for various interactions--also for unique J=0$^{+}$T=2 state.}
\begin{tabular}{ccccc}
\hline 
J$_{p}$J$_{n}$ & J=0 pairing & J=7 pairing & MBZE  & J=0 T=2\tabularnewline
\hline 
0 0 & 0.8660 & 0.6486 & 0.7878 & -0.5000\tabularnewline

2 2 & 0.2152 & 0.7143 & 0.5616 & 0.3737\tabularnewline

4 4 & 0.2887 & 0.1452 & 0.2208 & 0.5000\tabularnewline

6 6 & 0 .3469 & 0.0058 & 0.1234 & 0.6009\tabularnewline
 
\end{tabular} 
\end{center}
\vspace{10 pt}

Note that the even J 2-body matrix elements above have isospin T=1 and the odd ones have isospin T=0. The even states are antisymmetric and can occur for 2 neutrons,
2 protons and for a neutron and a proton The odd J states are symmetric
and can only exist for the proton neutron system $^{42}$Sc. Although
J=0 is the lowest there are also low lying J=1 and J=J$_{max}$ =7
two body matrix elements. This results in the fact
that in the ground state of $^{44}$Ti there is a high probability
that the 2 protons couple to J$_{p}$=2 and neutrons to J$_{n}$=2.
For MBZE{[}5{]} the amplitude is 0.5616 or 31.5\% probability. This
may be regarded as a precursor to works which show the importance
of L=2 couplings and of d bosons including IBA {[}1,2{]}.

We also show the wave functions for a J=0 pairing interaction and
likewise J=J$_{max}$=7. In the former case there is less (2,2) coupling
than MBZE. One might wonder why one does not get 100\% (0,0). This
is shown in the last column where the unique J=0 T=2 state is shown.
This is a double analog of a unique J=0 state of 4 neutrons i.e. $^{44}$Ca
and so the wave function is unique and it necessary has 25\% of the
(0,0) strength. Foe J=J$_{max}$ one gets more (2,2) coupling than
for MBZE.

\section{SENIORITY FOR PROTON NEUTRON SYSTEMS (J=0 PAIRING ): ALSO J$_{max}$
INTERACTION}

For the most part seniority considerations are generally applied to
systems of identical particles, and there are well documented works
including {[}6-7{]}. However there are examples where seniority is
relevant to systems of mixed neutrons and protons.

One example is the set of single j wave functions of $^{48}$Cr. In
the f$_{7/2}$ model space the valence nucleons are at mid-shell.
The quantity s=(-1)$^{x}$ with x=(v$_{p}$+v$_{n}$)/2 is a good
quantum number. Some states have s=+1 and others s=-1.This was shown
by Escudeors, Zamick and Bayman {[}5{]} and further work was done by
Neergaard {[}8{]} and Kingan et al. {[}9{]}.Note that neither v$_{p}$,v$_{n}$,
or v itself are good quantum numbers in the case-only s.

Another example is the pairing Hamiltonian of Edmond and Flowers{[}10{]}.
The states of mixed neutron-proton systems have quantum numbers (v,T,t)
where v is the seniority, T the total isospin and t the reduced isospin. The
latter is the isospin of nucleons not coupled to J=0. The energies
of the states for n nucleons is given by:

C$\{\frac{(n-v)(4j+8-n-v)}{4} -T(T+1) +t(t+1)\}$.

There are several selection rules for M1 transitions in this model, as
noted by Harper and Zamick{[}11{]}. The most interesting one is that
although seniority can change by 2 units one cannot change seniority
and reduced isospin at the same time. This is in contrast to the case
for identical particles where the M1 operator cannot change the seniority.
To explain this we note that the M1 operator must act on an np pair. If
the M1 operator acts on a J=0(T=1) pair it creates a J=1 (T=0) pair.
Since the new pair has T=0 it will not affect the reduced isospin. Alternately
if we act on a J=1 T=0 pair to create a J=0 T=1 pair, we note that
because the initial pair has T=0 it does not affect the reduced isospin.
On the other hand the seniority has been change by 2 units.

We look at the Edmond Flowers energy formula above{[}10{]} in some
detail using $^{96}$Cd as an example--2 proton holes and 2 neutron
holes in the g$_{9/2}$ shell. Note that there is no J in the formula.
This leads to high degeneracy. With appropriate C and scaling the
ground state energy is E=0 with seniority v=0. For v=2 we get states
at E=1 MeV with even J values 2,4,6 and 8. Then we get v=4 states
with J=10,12,14 and 16 at E=2.24 MeV. Note the big break in the spectrum
from J=8 and 10. Zamick{[}12{]} claims that with realistic interactions
there is also a mid-J break, although not so pronounced{[}10{]}. Using
the realistic interaction of Qi {[}13{]} the differences E(J)-E(J+2)
for J=0,2,4---16 are respectively 0.892,1.113, 1.041, 0.348, 1.681,
0.528, 0.128, and -0.216. We see that there is indeed a gap E(8)-E(10)
=1.681 MeV. This is much larger than the neighboring differences.
The concept of seniority splitting is behind this break. It deserves
further investigation.

With regards to identical particles Escuderos and Zamick {[}14{]}
noted that although seniority should not be good beyond j=7/2 there
were 2 unique v=4 states in the (g$^{9/2}$)$^{4}$configuration,
one with J=4$^{+}$ and the other with J=6$^{+}$ which remained eigenstates
even when seniority violating interactions were used{[}14{]}. See
further work by Y. Quian and C. Qi 15{]} and references therein.

Consider 2 holes in the g$_{9/2}$ shell. We define ithe J$_{max}$
interactions as being zero except when J=J$_{max}$ in which case
it is a negative constant. Zamick and Escuderos {[}16{]} compared
the spectrum of 2 proton holes and 2 neutron holes ( i.e. $^{96}$Cd)
for such an interaction with realistic CCGI {[}17{]} as well as one
with half E(J$_{max}$) and half E(0).Results are shown in Table 2.

\begin{center}

\captionof{table}{Spectra of yrast states with various interactions}
\begin{tabular}{cccccccccc}
\hline 
I & 0 & 2 & 4 & 6 & 8 & 10 & 12 & 14 & 16\tabularnewline
\hline 
CCGI & 0 & 1.081 & 2.112 & 2.888 & 3.230 & 4.882 & 5.339 & 5.403 & 5.224\tabularnewline

E(9) & 1.059 & 1.059 & 1.059 & 1.059 & 1.051 & 1.046 & 0.967 & 0.657 & 0.000\tabularnewline

E(0,9) & 0 & 1.274 & 1.858 & 2.393 & 2.512 & 3.214 & 3.135 & 2.847 & 2.168\tabularnewline
 
\end{tabular}
\end{center}
\vspace{10 pt}
We see that the CCGI interaction {[}17{]} gives a steady increase
in excitation energy with angular momentum except for J=16. The 16$^{+}$
state is correctly predicted to lie below the 14$^{+}$state. This
explains why the 16$^{+}$ state is isomeric.

With E(9) we see a terrible spectrum with for the most part the excitation
energy decreasing with angular momentum and having J$_{max}$=16
as the ground state. At first this might sound surprising, but the
J=16$^{+}$ wave function can be written as (pn)J=9 (pn)J=9 (antisymmetrized).
We can see how this feels the attraction of 2 nucleons with J=9.
Things are more reasonable if one takes half and half mixtures of
E(0) and E(9). 

Despite the bad spectrum with the E(9) interaction it turns out the
wave functions are quite good. Furthermore they are to an excellent
approximation proportional to unitary 9j coefficients. In some cases
the proportionality is exact {[}18,19,16{]}.

Just to give an example we compare the lowest 2 J=2$^{+}$states in
$^{96}$Cd--E(9) interaction results vs U9j. The U9j columns are suitably
normalized sets ( (jj)$^{9}$(jj)$^{J_{x}}$\textbar{} (jj)$^{J_{p}}$(jj)$^{J_{n}}$)$^{J=2}$.
In the first case J$_{x}$=9 and in the second case J$_{x}$=7.
We see that J$_{x}$ is an approximate quantum number.

As seen in Table 3 there is stunning agreement between the wave function
amplitudes obtained by a diagonalization with the E(9) interaction
and those from U9j. This leads to the question as to why there is
no more mixing of the 2 J$_{x}$states. Part of the reason lies in
the value of the U9j 

( (jj)$^{9}$(jj)$^{9}$\textbar{} (jj)$^{9}$(jj)$^{7}$)$^{J=2}$. We
note it is fairly large for j =3/2 but becomes increasingly small
with j. This lead us{[}19{]} to study the asymptotic value of this
U9j. Whereas the expression for this U9j is very complicated the asymptotic
value is simple. For large j the expression for this U9j is A j$^{m}$
e$^{(-\alpha j)}$with m=$\frac{3}{2}$ and $\alpha$= 4 ln(2). {[}18{]}We see
that it deceases almost exponentially with j. To obtain this result
we used the Stirling approximation: ln (n!)= n ln(n)-n +ln($\sqrt{2n\pi}$).
\begin{center}
\captionof{table}{Comparison of the wave functions of the 2 lowest 2$^{+}$
states: E(9) vs. U9j}
\begin{tabular}{ccccc}
\hline 
J$_{p}$,J$_{n}$ & E(9) & U9j(J$_{x}$=9) & E(9) & U9-j (J$_{x}$=7)\tabularnewline
\hline 
E{*} & 1.069 & . & 3.059 & .\tabularnewline

0,2 & 0.5334 & 0.5338 & 0.1349 & 0.1351\tabularnewline
 
2,2 & -0.4707 & -0.4708 & 0.5569 & 0.5567\tabularnewline

2,4 & 0.3025 & 0.3035 & 0.3188 & 0.3189\tabularnewline

4,4 & -0.1388 & -0.1390 & 0.6300 & 0.6299\tabularnewline

4,6 & 0.0531 & 0.0531 & 0.1320 & 0.1320\tabularnewline

6,6 & -0.0137 & -0.0138 & 0.1350 & 0.1350\tabularnewline

6,8 & 0.0025 & 0.0025 & 0.0114 & 0.0014\tabularnewline

8,8 & -0.0003 & -0.0003 & 0.0052 & 0.0052\tabularnewline
 
\end{tabular}
    
\end{center}

\section{CALCULATED INTER-BAND B(E2)'s IN THE P-F SHELL}
Previous studies of even-even Ti isotopes showed reasonably strong
B(E2)'s in the yrast band--J=0$_{1}$ to 2$_{1}$, 2$_{1}$ to 4$_{1}$,
e.t.c..{[}20{]}. In this work we study transitions from states in the
yrast band to a second group of states: 1$_{1}$,2$_{2},3_{1}$ ,4$_{2}$
,5$_{1}$ i.e. second excited states of even J and lowest states of
odd J. We use the shell model code NushellX {[}21,22{]}. We make comparisons
with the rotational model as described by Bohr and Mottelson{[}23{]}

In Table 4 we show the yrast transitions. The largest ones are
from J=0 to J=2. The values for the Ti isotopes with A=44,46,48 and
50 in e$^{2}$ fm$^{4}$ are respectively 526, 624, 521 ad 502 e$^{2}$fm$^{4}$
There is a large increase in $^{48}$Cr ,1254 e$^{2}$fm$^{4}$. It
was noted in ref{[}20{]} that in both the rotational model and vibrational
model {[}23{]}, the J$\rightarrow J+2$ B(E2)'s increase with J ,but
as seen in table 1 in the shell mode they decrease with J.

We next make a comparison of the behavior in the Ti isotopes with
what occurs in more deformed nuclei. It is convenient to choose the
work of Clement et al. {[}24{]} on $^{98}$Sr because they show several
measured B(E2)'s between states in the yrast band and those in the
next band. The comparison is somewhat hybrid because we are listing
experimental results for Sr and theoretical results for Ti. The B(E2's)
in Weisskopf units (WU) are 19.4 in $^{46}$Ti and 95.5 in $^{98}$Sr.
This shows that the latter nucleus is indeed more strongly deformed
than any of the Ti isotopes.

In their Table 4 Clement et al.{[}24{]} show reduced matrix elements.
In our Table 5 we show rather the ratio of a given B(E2) to the intraband
0(1)$\longmapsto2(1)$ B(E2). The ratio of this transition to 2(1)$\longmapsto2(2)$
in $^{98}$Sr is quite small whereas for $^{44}$Ti and $^{46}$Ti
the values are 0.2909 and 0.1694 respectively. A ratio close to 0.2
is also found for 0(1)$\rightarrow$2(2) in$^{48}$Ti.

In Table 6 we show B(E2)'s within the second group for $^{46}$Ti.
They are in general much larger than the interband transitions between the 2 groups.
\vspace{20 pt}

\begin{centering}

\newpage

\captionof{table}{Calculated yrast B(E2) J$\rightarrow$(J+2) e$^{2}$fm$^{4}$}
\begin{tabular}{cccccc}
\hline 
J  & $^{44}$Ti  & $^{46}$Ti & $^{48}$Ti & $^{50}$Ti & $^{48}$Cr \tabularnewline
\hline 
0  & 526  & 624  & 521 & 502&1254 \tabularnewline
 
2  & 246  & 286  & 269 &176&609\tabularnewline
 
4  & 155  & 228  & 87.8&67.9&487\tabularnewline
 
6  & 94.7  & 190  & 102&.426&403\tabularnewline

8  & 114  & 134  & 68.2&57.5&261\tabularnewline

10  & 64.1  & 49.8  & 28.8&56.5&194\tabularnewline

12  &  & 45.0  & 5.26 &13.6&148\tabularnewline

14  &  & 0.062  & 7.25 &22&71.3\tabularnewline

\end{tabular}

\end{centering}

\begin{centering}

\captionof{table}{Ratio B(E2)/B(E2) 0(1)$\rightarrow$2(1)}
\begin{tabular}{ccccccc}
\hline 
Ji$\rightarrow$Jf  & $^{98}$Sr  & $^{44}$Ti  & $^{46}$Ti  & $^{48}$Ti& $^{50}$Ti& $^{48}$Cr\tabularnewline
\hline 
0(1)$\rightarrow$2(2)  & 0.00799  & 0.05380  & 0.006458  & 0.1902& 0.000142& 0.00246\tabularnewline

2(1)$\rightarrow$0(2)  & 0.02556  & 0.0113  & 0.0208  & 0.0195&0.00219& 0.0221\tabularnewline

2(1)$\rightarrow$2(2)  & 0.000767{*}  & 0.2909  & 0.1694  & 0.0845& 0.00703&0.0451\tabularnewline

4(1)$\rightarrow$2(2)  & 0.004603  & 0.02567  & 0.01651  & 0.00263& 0.00365&0.00606\tabularnewline

2(1)$\rightarrow$3(1)  &  & 0.0123  & 0.009295  & 0.0595& 0.000448& 0.0151\tabularnewline
 
4(1)$\rightarrow$3(1)  &  & 0.04297  & 0.006490  & 0.04626& 0.0307& 0.00177\tabularnewline

\end{tabular}

\begin{table}[H]
\captionof{table}{$^{46}$TI B(E2) e$^{2}$fm$^{4}$ in the second group}
\begin{tabular}{cccccc}
\hline 
 & J-2(2)  & J-1(1)  & J(2)  & J+1(1)  & J+2(2) \tabularnewline
\hline 
0(2)  & x  & x  & x  & x  & 1.61E+02 \tabularnewline

2(2)  & 3.21E+01  & 1.26E+01  & 1.76E+01  & 1.29E+01  & 5.06E+01 \tabularnewline

4(2)  & 2.81E+01  & 2.05E+01  & 4.14E+00  & 7.19E+00  & 1.54E+00 \tabularnewline

6(2)  & 1.07E+00  & 3.49E+01  & 3.79E+00  & 4.98E+00  & 1.07E+02 \tabularnewline

8(2)  & 8.16E+01  & 1.69E+01  & 2.00E+00  & 3.81E+00  & 7.93E+01 \tabularnewline

10(2)  & 6.42E+01  & 2.16E+01  & 1.22E+01  & 3.41E+01  & 1.39E+01 \tabularnewline
 
12(2)  & 1.17E+01  & 2.22E+00  & 2.99E+00  & 2.05E+01  & 1.13E-01 \tabularnewline

14(2)  & 9.73E-02  & 9.85E-01  & 1.16E+02  & 4.19E-01  & 8.59E-01 \tabularnewline

16(2)  & 7.55E-01  & 1.27E+00  & 1.59E+02  & x  & x \tabularnewline

\end{tabular}
\end{table}
\end{centering}

\begin{centering}

\captionof{table}{Electric quadrupole moments (e fm$^2$)}
\begin{tabular}{cccccc }
\hline
     &$^{44}$Ti&$^{46}$Ti&$^{48}$Ti&$^{50}$Ti& $^{48}$Cr \\
     \hline
    2(1) &6.01&-13.6&-14.5&6.53&-30.8\\
    2(2)&-0.89&7.1&5.02&13.3&21.9\\
\end{tabular}

\end{centering}

Note that for $^{46}$Ti,$^{48}$Ti and $^{48}$Cr the quadrupole moments of the 2(1) states are negative and
those of the 2(2) states are positive. In the rotational model the first group would be prolate and
the second group would be oblate. Indeed, the quadrupole moments of J=2+ for a K=2 band are
equal and opposite of those of a K=0 band.

In closing we note that the main point of this work is that inter-band B(E2)'s show variations as
one goes from one isotope to another and as one goes from less deformed to more strongly
deformed nuclei. The behaviors are not as well understood as those for yrast transitions. They
deserve to be studied more both experimentally and theoretically. We note that the calculated intra-band transitions in the first group are large and likewise for the second group. The inter-
band transitions between the 2 groups are very small and it is hard to see a trend with increasing neutron number.

\section{MATRIX STUDIES}

In almost anything we do, be it shell model, IBM, or anything else, somewhere we are diagonalizing a matrix. So we thought it would
be a good idea to study the properties of matrices in general, somewhat
divorced from specific experiments. To this end with Arun Kingan {[}25,26{]}
we started with the matrix shown in Figure 1 (a) (although ours were
11 by 11).The diagonal is nE and there are constant interaction matrix
elements are v only next to the diagonal. We defined 2 possible transition
operators \textless{}n T$_{A}$(n+1)\textgreater{} =1 and \textless{}n
T$_{B}$ (n+1)\textgreater{} = $\sqrt{(n+1)}$, with all other transition
elements being zero. With the B choice we found a near exponential
decrease of transition strength with excitation energy for all v.
We found a similar behavior with the A choice for small v but for
large v we find 2 exponential behaviours- one for even n and one for
odd n{[}26{]}

Here for simplicity we take the B choice and consider only the case
v=1. As seen in Fig.2 we do indeed get a simple exponential behavior
for the tridiagonal case-on a log plot an almost straight line with
a negative slope. We then consider a new pentadiagonal matrix(Figure 1 (b)).

\begin{center}

\captionof{figure}{ Study of Tridiagonal (a) and Pentadiagonal Matricies (b).}
 (a) \hspace{95 pt} (b)\\
$ \begin{bmatrix}
0 & v & 0 & 0 & 0 \\
v & E & v & 0 & 0 \\
0 & v & 2E & v & 0 \\
0 & 0 & v & 3E & v \\
0 & 0 & 0 & v & 4E
\end{bmatrix}  $
\hspace{10 pt}
$\begin{bmatrix}
0 & v & v & 0 & 0 \\
v & E & v & v & 0 \\
v & v & 2E & v & v \\
0 & v & v & 3E & v \\
0 & 0 & v & v & 4E 
\end{bmatrix}$

\vspace{10 pt}

\begin{figure}[h]

  \begin{tabular}[b]{c}
    \includegraphics[width=.3\linewidth]{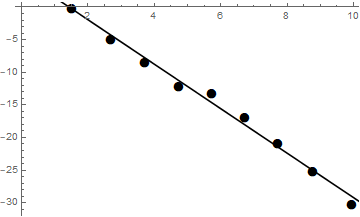} \\
    \small (a) Tridiagonal
  \end{tabular} \qquad
  \begin{tabular}[b]{c}
    \includegraphics[width=.3\linewidth]{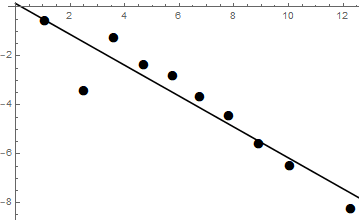} \\
    \small (b) Pentadiagonal
  \end{tabular}
  \caption{Transition strength versus excitation energy}
\end{figure}

\end{center}
Figure 2 (a) (Tridiagonal) does indeed display a simple near exponential
fall off of the transition strength. Although we just said we are
divorcing ourselves from experiment we note that exponential fall
offs have been frequently observed or calculated e.g. in gamma cascades
following neutron capture {[}27,28,29{]}; also in calculated magnetic
dipole strength.The ground state wavefunction has an interesting structure in
the weak coupling limit:  a$_{0}$=1, a$_{1}$ = $-\frac{v}{E}$, a$_{2}$
= $\frac{v^2}{E^22!}$, a$_{3}$ =  $-\frac{v^3}{E^33!}$, ..., a$_{n}$
=  $\frac{(-v)^n}{E^n n!}$. This coincides with the Taylor
series for the exponential function e$^{-\frac{v}{E}}$. We can understand
the factorials in the denominator the fact that in n\textquoteright th
order perturbation theory the energy denominators are (E0-E1) (E0-E2)
........ (E0-En), which ,with our choice of a matrix is $(-E)^nn!$.

In Fig 2 (b) (Pentadiagonal) we see more structure, one can indeed draw a
straight line through many of the points but at low energies there
is a dip. These studies are in their infancy and there is still much to
learn.

\end{document}